\documentclass[useAMS,usenatbib]{mn2e}

\usepackage{graphicx,bm,amsmath,ulem}

\newcommand{\Msol}{\mbox{$\mathrm{M}_{\sun}$}}
\newcommand{\Msun}{\mbox{$\mathrm{M}_{\sun}$}}

\newcommand{\Mpc}{\mbox{$\mathrm{Mpc}$}}

\newcommand{\tr}{\mbox{$\mathrm{tr}$}}

\title[Weak lensing analysis of RXC J2248.7-4431]{Weak lensing analysis of RXC J2248.7-4431}
\author[D. Gruen et al.]
{\parbox{\textwidth}{D. Gruen$^{1,2}$\thanks{E-mail:
dgruen@usm.uni-muenchen.de},
F. Brimioulle$^{1,2}$,
S. Seitz$^{1,2}$,
C.-H. Lee$^{1,2}$,
J. Young$^{3}$,
J. Koppenhoefer$^{1,2}$,
T. Eichner$^{1,2}$,
A. Riffeser$^{1,2}$,
V. Vikram$^{4}$,
T. Weidinger$^{1,2}$ and
A. Zenteno$^{1}$}\vspace{0.4cm}\\
\parbox{\textwidth}{$^{1}$University Observatory Munich, Scheinerstrasse 1, D-81679 Munich, Germany\\
$^{2}$Max Planck Institute for Extraterrestrial Physics, Giessenbachstrasse, D-85748 Garching, Germany\\
$^{3}$Ohio State University, Department of Physics, 191 West Woodruff Avenue, Columbus, OH 43210, USA\\
$^{4}$Department of Physics and Astronomy, University of Pennsylvania, 209 South 33rd Street, Philadelphia, PA 19104, USA
}}
\begin{document}

\date{}

\pagerange{\pageref{firstpage}--\pageref{lastpage}} \pubyear{2013}

\maketitle

\label{firstpage}

\begin{abstract}
We present a weak lensing analysis of the cluster of galaxies RXC J2248.7-4431, a massive system at $z=0.3475$ with prominent strong lensing features covered by the Cluster Lensing
And Supernova survey with Hubble (CLASH). Based on \textit{UBVRIZ} imaging from the Wide-Field Imager camera at the MPG/ESO 2.2-m telescope, we measure photometric redshifts and shapes of background galaxies. The cluster is detected as a mass peak at $5\sigma$ significance. Its density can be parametrized as a Navarro-Frenk-White (NFW) profile with two free parameters, the mass $M_{200m}=33.1^{+9.6}_{-6.8}\times 10^{14}\Msol$ and concentration $c_{200m}=2.6^{+1.5}_{-1.0}$. We discover a second cluster inside the field of view at a photometric redshift of $z\approx0.6$, with an NFW mass of $M_{200m}=4.0^{+3.7}_{-2.6}\times10^{14}\Msol$.
\end{abstract}


\section{Introduction}

\begin{figure}
\centering
\includegraphics[width=0.48\textwidth]{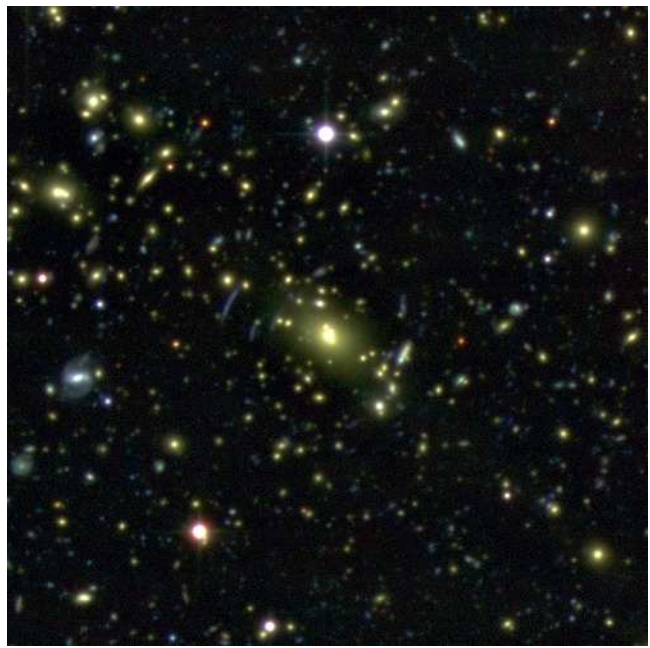}
\caption{BRI colour image of the central 200 $\times$ 200~arcsec$^2$ of RXC J2248.7-4431. Clearly visible are the brightest cluster galaxy and several gravitational arcs, along with a high concentration of yellow cluster member galaxies.}
\label{fig:colorimage}
\end{figure}

Clusters of galaxies, such as RXC J2248.7-4431 (cf. Fig.~\ref{fig:colorimage}) studied in this work, are the most massive gravitationally bound objects that have formed in the Universe to date. 
What makes them interesting for cosmology is that they lie at an intersection of the two potentially most important unresolved questions: their mass content is dominated by dark matter - and their formation and evolution is strongly influenced by the interplay of matter density and dark energy. For this reason, studying clusters of galaxies is also a powerful probe of cosmological parameters and models \citep[e.g.][]{1998ApJ...508..483W,2001ApJ...560L.111H,PhysRevLett.88.231301}.

Perhaps the most important property of a cluster of galaxies with respect to cosmology is its virial mass, which can be determined in several different ways. The number and the velocity dispersion of cluster member galaxies are related to the total mass of the system. One can use larger fractions of the overall mass as a proxy by observing the hot intracluster gas, which emits X-ray radiation \citep[e.g.][]{2004AuA...425..367B,2009MNRAS.397..577S,2010arXiv1007.1916P} and changes the cosmic microwave background spectrum due to inverse Compton scattering \citep[Sunyaev-Zel'dovich, SZ, effect; cf.][]{1972CoASP...4..173S,2001ApJ...553..545H,2003PhRvD..68h3506B}. The latter observables can be related to mass by astrophysical modelling or self-calibration \citep{2003PhRvD..67h1304H,2004ApJ...613...41M}. 

Despite this, however, weak gravitational lensing is a valuable ingredient since the tangential alignment of background galaxy images is directly proportional to overdensity of all cluster matter - luminous or dark - alike and is insensitive to the astrophysical state of the cluster. This allows for unbiased mass measurements of single clusters and improved calibration of other mass-observable relations \citep[e.g.][]{2002MNRAS.334L..11A,2010ApJ...721..875O,2011ApJ...726...48H}. Furthermore, gravitational lensing straightforwardly allows the probing of additional properties of the density profiles of dark matter haloes, such as their concentrations, for which predictions in a cosmological model can be made and from which additional constraints can be drawn \citep[see for instance the Cluster Lensing And Supernova survey with \textit{Hubble} (CLASH), in which RXC J2248.7-4431 is also observed; cf.][]{2012ApJS..199...25P}.

In this work we analyse the weak lensing effect of the cluster RXC J2248.7-4431 based on background galaxy shapes and photometric redshifts measured from \textit{UBVRIZ} multiband imaging by the Wide-Field Imager (WFI) on the 2.2-m MPG/ESO telescope at La Silla. In Section~2, we introduce the basic properties of the data used and our data reduction. Section~3 summarizes previous optical, X-ray and SZ observations of RXC J2248.7-4431. We give an overview of the photometric analysis, including photometric redshifts, cluster member photometry and morphology, in Section~4. Technical aspects of our weak lensing measurements are discussed in Section~5. The analysis of the weak lensing effect of RXC J2248.7-4431 is presented in Section~6. Results for a second cluster found in the field of view at $z\approx0.6$ are shown in Section~7. We summarise our results in Section~8.

All numerical values given in this work are calculated for cosmological parameters $H_0=72$~km s$^{-1}$ Mpc$^{-1}$ and $\Omega_m=1-\Omega_{\Lambda}=0.27$. Where applicable, measurements from the literature have been converted to this cosmology as well. We denote the radii of spheres around the cluster centre with fixed overdensity as $r_{\Delta \rm{m}}$ and $r_{\Delta \rm{c}}$, where $\Delta$ is the overdensity factor of the sphere with respect to the mean matter density $\rho_{\rm m}$ or critical density $\rho_{\rm c}$ of the universe at the cluster redshift. Masses inside these spheres are labelled and defined correspondingly as $M_{\Delta \rm{m}}=\Delta\cdot\frac{4\pi}{3}r_{\Delta \rm{m}}^3\cdot\rho_{\rm m}$ and $M_{\Delta \rm{c}}=\Delta\cdot\frac{4\pi}{3}r_{\Delta \rm{c}}^3\cdot\rho_{\rm c}$.

\section{Observations and Data reduction}

This analysis is based on observations made with the WFI on the 2.2-m MPG/ESO telescope at La Silla. The sensitivity of the instrument and the filters available spans all optical wavelengths and with its field of view of 33 $\times$ 33~arcmin$^2$ it is well suited for weak lensing cluster analyses with photometric redshifts.

\begin{figure}
\centering
\includegraphics[width=0.48\textwidth]{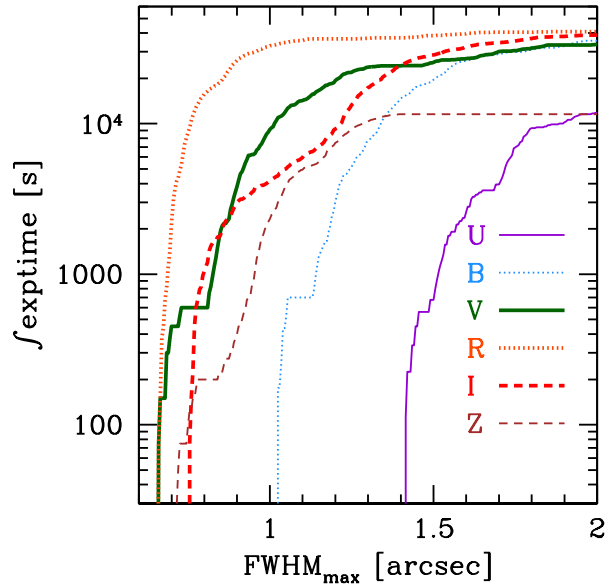}
\caption{Cumulative exposure time as a function of limiting seeing in WFI frames of RXC J2248.7-4431 used in this work. Bold lines show the bands used for shape measurement, where we have applied seeing FWHM cuts at 0.9~arcsec (1.0~arcsec) in \textit{R} (\textit{V} and \textit{I}) band, respectively.}
\label{fig:exptime}
\end{figure}

The observations in \textit{U} (\#877), \textit{B} (\#842), \textit{V} (\#843), \textit{R} (\#844), \textit{I} (\#879) and \textit{Z} (\#846) band\footnote{\texttt{https://www.eso.org/lasilla/instruments/wfi/inst/filters/}} used in this work were taken in the years 2009-2010. Details of integration time as a function of limiting point-spread function (PSF) full-width at half-maximum (FWHM) are shown in Fig.~\ref{fig:exptime}. The excellent depth and seeing in \textit{R} band (with over 9 hours of exposure time at sub-arcsecond seeing, yielding a coadded image with 0.8~arcsec PSF FWHM and limiting magnitude of $m_{R,\mathrm{Vega},\mathrm{lim}}=26.7$ for a $5\sigma$ detection inside a 1~arcsec diameter aperture) make it our primary lensing band, but also in \textit{V} (0.9~arcsec seeing with $m_{V,\mathrm{Vega},\mathrm{lim}}=26.4$) and \textit{I} band (0.9~arcsec PSF FWHM with $m_{I,\mathrm{Vega},\mathrm{lim}}=24.4$) shape measurements of background galaxies are feasible. 

We perform de-biasing, flat fielding, masking of bad pixels in all bands and fringe pattern correction in the \textit{I} and \textit{Z} band using the \textsc{Astro-WISE}\footnote{\texttt{http://www.astro-wise.org/}} pipeline \citep{2007ASPC..376..491V}. Background subtraction, final astrometry and co-addition of suitable frames is done externally using \textsc{SCAMP}\footnote{\texttt{http://www.astromatic.net/software/scamp}} \citep{2006ASPC..351..112B} and \textsc{SWarp}\footnote{\texttt{http://www.astromatic.net/software/swarp}} \citep{2002ASPC..281..228B}. The central part of a colour image based on these co-added frames is shown in Fig.~\ref{fig:colorimage}.

For photometry, we take observations of standard star fields in \textit{B}, \textit{V} and \textit{R} to fit zero-points for each CCD individually and extinction coefficients globally. From nights which we confirm to be photometric according to their standard star measured zero-points we build a photometric \textit{B}, \textit{V} and \textit{R} stack. We build deeper stacks including frames from all available nights with relative photometry fitted by \textsc{SCAMP} and match the magnitudes measured from these stacks to the photometric ones in order to find zero-points for the deep stacks. Alternatively, fixing the \textit{R} band zero-point, we find \textit{U}, \textit{B}, \textit{V}, \textit{I} and \textit{Z} zero-points by minimizing residuals in colour-colour diagrams with respect to the stellar library of \citet{1998PASP..110..863P}. The stellar locus zero-points found in this way for \textit{B} and \textit{V} agree within 0.02~mag with the photometric zero-points found by matching the two versions of the stacks, confirming that our photometric pipeline provides consistent results.

\section{Previous work on RXC J2248.7-4431}

The cluster RXC J2248.7-4431 studied in this work is also known as Abell S1063 or MACS 2248-4431. Here we give an overview of all literature on the cluster published previously to this work, including detections and redshift estimates, X-ray observations and SZ measurement. We discuss the more detailed findings of \citet{2012AJ....144...79G} and compare them to our own analyses in Section~\ref{sec:gomez}.

RXC J2248.7-4431 was first listed by \citet{1989ApJS...70....1A} with a background-corrected galaxy count of 74. It was independently detected by the \textit{ROSAT} All-Sky Survey \citep{1999ApJ...514..148D}, who quote a redshift estimate of $z=0.252$, which was, however, only weakly determined based on the \citet{1989ApJS...70....1A} distance class. \citet{2002ApJS..140..239C} give its redshift as $z=0.1495$ based on private communications with Andernach (private communication). \citet{2004AuA...425..367B} finally quote a spectroscopic redshift of $z_{\mathrm{cl}}=0.3475$, which is what we adopt for this work. This is confirmed by spectroscopy of 51 cluster members with a mean redshift of $z=0.3461^{+0.0010}_{-0.0011}$ \citep{2012AJ....144...79G}.

\citet{2012AJ....144...79G} find the spectroscopic velocity dispersion of to be $\sigma_v=1660^{+230}_{-150}$km s$^{-1}$, which corresponds to a mass of $M_{200c}=42^{+17}_{-9}\times10^{14}\Msun$ according to the relation of \citet{2008ApJ...672..122E}.

\citet{2002ApJS..140..239C} quote an X-ray temperature of 7.823~keV from the \textit{ROSAT} All-Sky Survey (RASS). \citet{2008ApJS..174..117M} give the X-ray temperature within $R_{500}$ as $11.1^{+0.8}_{-0.9}$keV and parametrize the cluster profile as slightly elliptical with $1-b/a=0.2$ based on \textit{Chandra} ACIS-I data. Both values are based, however, on earlier erroneous cluster redshifts ($z=0.1495$ for \citealt{2002ApJS..140..239C} and $z=0.252$ for \citealt{2008ApJS..174..117M}). The \citet{2011AuA...536A..11P} quote an X-ray mass based on their follow-up with \textit{XMM-Newton} of $M_{500c}=(12.25\pm0.21)\times10^{14}\Msun$ which they use to calibrate their SZ MOE. \citet{2011MNRAS.418.1089C} determine $M_{2500c}=(5.3\pm2.6)\times10^{14}\Msol$ from \textit{Chandra} X-ray data and \citet{2012AJ....144...79G} give a consistent value from independent data reduction of the same data of $M_{2500c}=(6.0\pm1.6)\times10^{14}\Msol$.

\citet{2010ApJ...716.1118P} parametrize the SZ profile measured with the South Pole Telescope (SPT) with a $\beta$ parameter of $(0.86\pm0.02)$~arcmin at a scaling of $\Delta T\approx 1$~mK. \citet{2011ApJ...738..139W} quote an SZ signal-to-noise ratio (S/N) of 20.7 and a SZ mass of $M_{200m}=(28.2\pm_{\rm stat}3.6\pm_{\rm sys}9.3)\times10^{14}\Msun$. The \citet{2011AuA...536A...8P} detect the SZ effect of RXC J2248.7-4431 at $13.93\sigma$ significance. From the \textit{Planck} SZ observable and scaling relation \citep{2011AuA...536A..11P} the SZ mass is $M_{500c}=(11.5\pm_{\rm stat}2.6\pm_{\rm sys}0.5)\times 10^{14}\Msun$.

\citet{2009AuA...499..357G} and \citet{2010ApJ...716.1118P} show X-ray and SZ imaging of the cluster, respectively. The cluster is also covered by the CLASH project \citep{2012ApJS..199...25P,annamonna}.

\section{Photometric analysis}
\subsection{Photometric Redshifts}

The multicolour catalogue creation and photometric redshift estimation follow the procedure described in \citet{fabrice}. Here we only give a brief overview. We convolve all data with a Gaussian kernel to match the seeing to the band where it is largest (in this case the \textit{U} band). This equalization of the PSF is required for reliable aperture colours in the different filters. We then run \textsc{SExtractor}\footnote{\texttt{http://www.astromatic.net/software/sextractor}} in dual-image mode to extract fluxes, including weight images and masks of bad areas in the detection and extraction frame. We detect the objects on the unconvolved \textit{R} band with a S/N threshold of $2\sigma$ on at least four contiguous pixels. Flux and magnitude information is extracted from the convolved images. We correct for extinction and zero-point accuracies by comparing the stellar locus in colour-colour diagrams to the stellar library of \citet{1998PASP..110..863P}. We then use the photometric template-fitting algorithm of \citet{2001defi.conf...96B} to estimate photometric redshifts.

\begin{figure}
\centering
\includegraphics[width=0.48\textwidth]{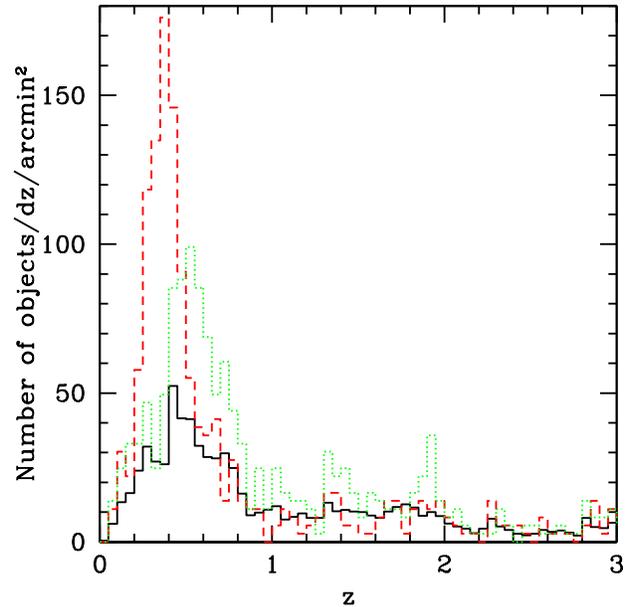}
\caption{Three-dimensional galaxy density as a function of photometric redshift in the WFI field (black, solid lines) and inside a circle of radius 100~arcsec around the brightest cluster galaxy of RXC J2248.7-4431 (red, dashed lines). The cluster at a redshift of $z_{\mathrm{cl}}=0.3475$ \citep{2004AuA...425..367B} can be clearly seen as a peak in this smaller field. The green, dotted line shows objects within a radius of 100~arcsec around the BCG of the $z\approx0.6$ cluster described in Section~\ref{sec:secondcluster}.}
\label{fig:photoz}
\end{figure}

The density of galaxies in redshift space is shown in Fig.~\ref{fig:photoz}. The cluster is clearly visible as a redshift density peak around the spectroscopic $z_{\mathrm{cl}}=0.3475$, which is also true for the second cluster at a mean photometric redshift of $z\approx0.6$, described in detail in Section~\ref{sec:secondcluster}.

\begin{figure}
\centering
\includegraphics[width=0.45\textwidth]{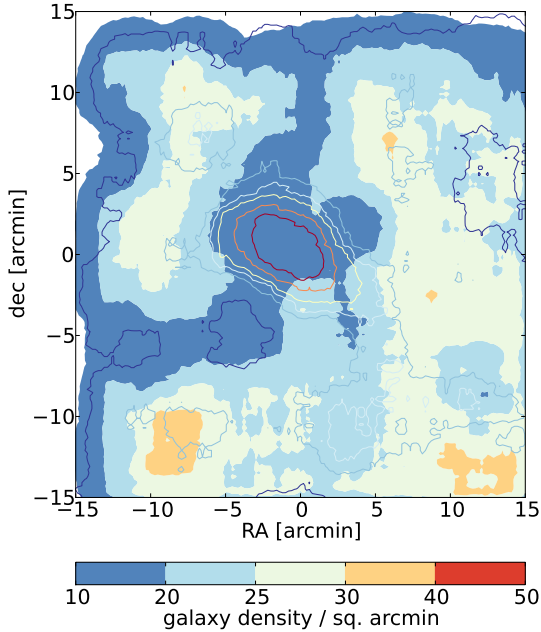}
\caption{Density of galaxies in our photometric redshift catalogue in a map centred on the BCG position, averaged in angular boxes of 4$\times$4~arcmin$^2$ size. Contour lines show density per arcmin$^2$ per redshift interval for sources with $|z-z_{\mathrm{cl}}|<0.04(1+z_{\mathrm{cl}})$. The background colour shows the overall density of sources above the redshift cut of equation~(\ref{eqn:zcut}). The depletion of background galaxies at the cluster position is an indication that our background sample is pure. The high background density observed in the lower left of the map is due to the system discussed in Section~\ref{sec:secondcluster}.}
\label{fig:zdens}
\end{figure}

In our weak lensing analysis we only consider background galaxies with a minimum photometric redshift of
\begin{equation}
z_{\mathrm{source}} > 1.1 z_{\mathrm{cl}} +0.15 \approx 0.53\ .
\label{eqn:zcut}
\end{equation}
We include objects up to a maximum photometric redshift of $z=4$. We verify that this excludes redshift mismeasurement of cluster galaxies into the background sample by means of redshift density maps at the cluster redshift and for objects selected by the background cut of equation~(\ref{eqn:zcut}) in Figure~\ref{fig:zdens}. Indeed, the cluster is seen as a highly significant overdensity of objects around $z_{\mathrm{cl}}$, while the density of background sources near the cluster is smaller than in the field. This is expected owing to the area covered by cluster light and indicates that background contamination is not a significant issue, as expected for the red cluster galaxies whose redshift can be determined photometrically to good accuracy.

\subsection{Cluster member SEDs}

We investigate photometric properties and spectral energy distribution (SED) types of the cluster member galaxies of RXC J2248.7-4431 and the second cluster at $z\approx0.6$ (cf. Section~\ref{sec:secondcluster}).

The brightest cluster galaxy (BCG) of RXC J2248.7-4431 has a photometric redshift of $0.4$ and an absolute magnitude of $M_{R,\mathrm{Vega}}=-24.81$. Its counterpart in the $z\approx0.6$ cluster has a photometric redshift of $0.66$ and an absolute magnitude of $M_{R,\mathrm{Vega}}=-25.1$. 

\begin{figure}
\centering
\includegraphics[width=0.48\textwidth]{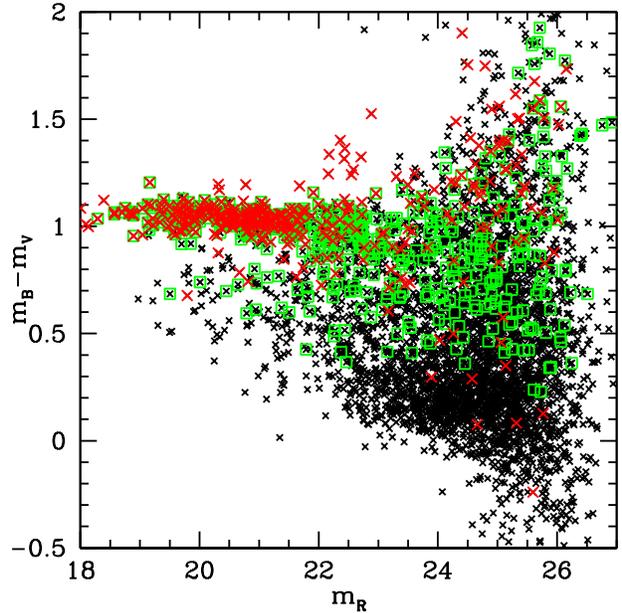}
\caption{Colour-magnitude diagram of galaxies within 5~arcmin projected separation from the BCG, using \textit{B}-\textit{V} aperture colours and \textit{R} band \textsc{MAG\_AUTO} magnitudes and showing all objects (black, small crosses), objects with red spectral energy distribution types \citep[cf.][]{2005ApJ...631..126D,fabrice} (red, large crosses) and marking objects in a redshift slice $|z-z_{\rm cl}|\leq0.04(1+z_{\mathrm{cl}})$ around the cluster by a green open square.}
\label{fig:rs}
\end{figure}

\begin{figure*}
\centering
\includegraphics[width=0.48\textwidth]{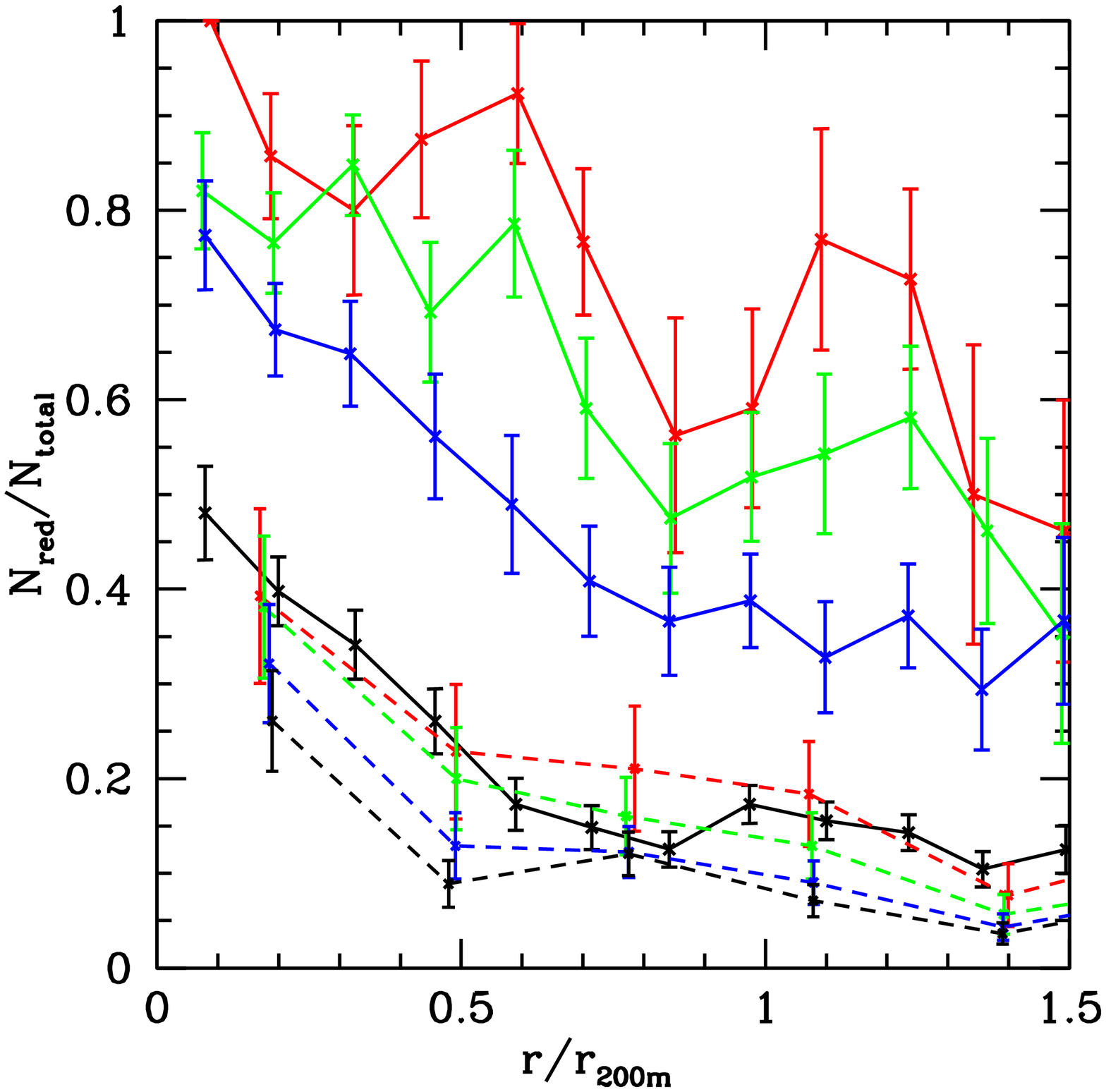}
\includegraphics[width=0.48\textwidth]{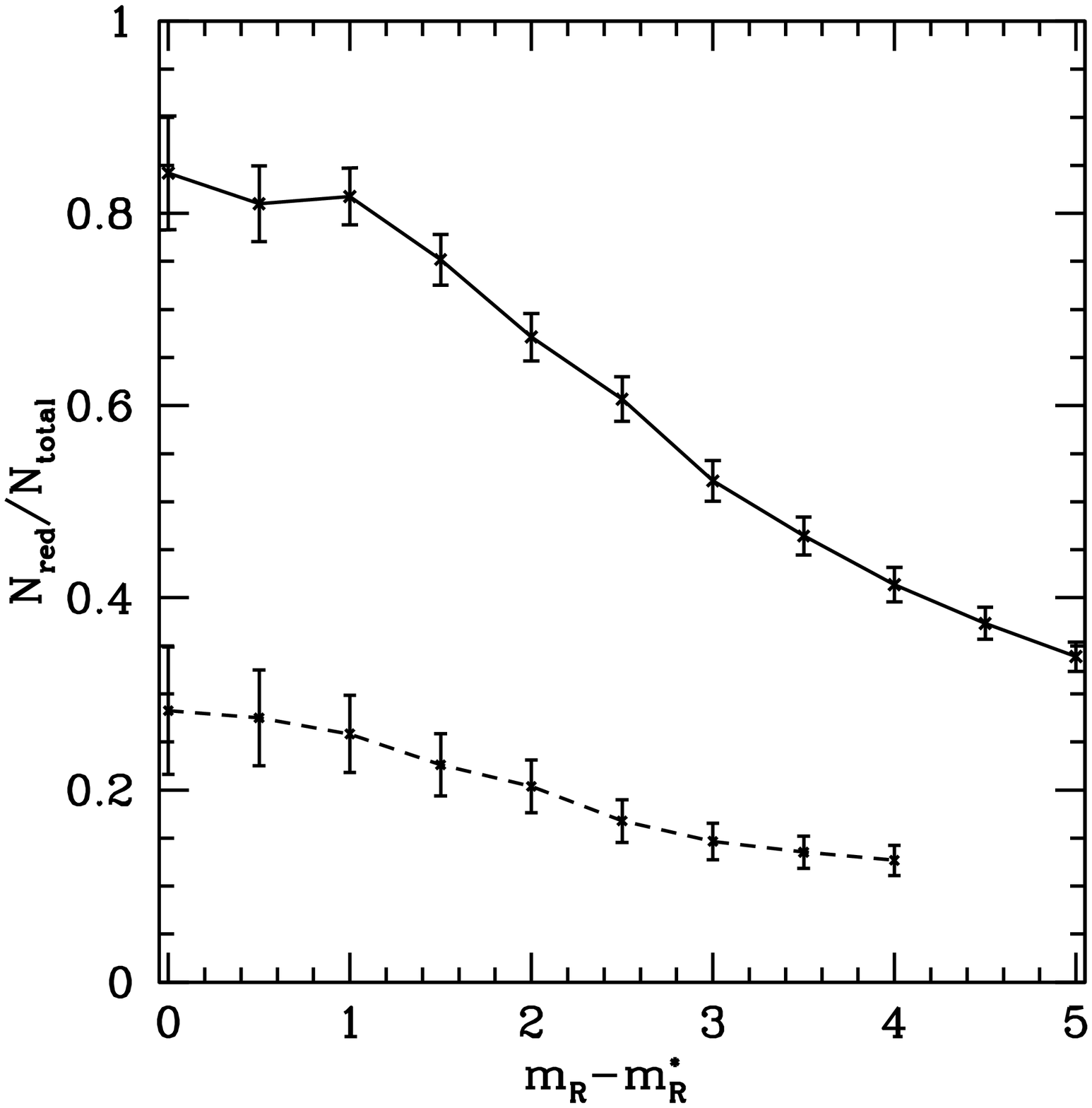}
\caption{Fraction of galaxies having a red SED type in the redshift slice and projected vicinity of RXC J2248.7-4431 (solid lines) and the second cluster at $z\approx0.6$ (dashed lines). The left panel shows the red fraction as a function of radius from the BCG in units of $r_{200m}$ of the respective system with magnitude limits at $m^{\star}_R+1$ (red), $m^{\star}_R+2$ (green), $m^{\star}_R+3$ (blue) and our detection limit (black). The right panel shows the red fraction inside $r_{200m}$ of the respective system as a function of magnitude limit, which reaches the overall detection limit at $m_{R}^{\rm lim}-m_R^{\star}\approx4$ for the $z\approx0.6$ system. Error bars are calculated from the Poissonian noise in the number counts of red and blue galaxies in each bin.}
\label{fig:bo}
\end{figure*}

Cluster members galaxies are preferentially of early type and, at a given redshift, lie on a relatively tight sequence in colour-magnitude space, the red sequence. We plot \textit{B}-\textit{V} aperture colours, which are particularly indicative of the D4000 break, against \textit{R} band \textsc{MAG\_AUTO} magnitudes in Fig.~\ref{fig:rs}.
The red sequence is clearly detected and dominant among cluster members. We measure the fraction of red galaxies on a sample of cluster members selected by their photometric redshift $|z-z_{\rm cl}|\leq0.04(1+z_{\mathrm{cl}})$ for both RXC J2248.7-4431 and the second cluster found at $z\approx0.6$. Fig.~\ref{fig:bo} shows the fraction of these galaxies which are classified photometrically as having a red SED \citep[cf.][]{2005ApJ...631..126D,fabrice}, $N_{\mathrm{red}}/N_{\mathrm{total}}$, as a function of projected separation from the cluster in units of its $r_{200m}$ according to our best-fitting Navarro-Frenk-White (NFW) models (see Sections \ref{sec:nfw} and \ref{sec:secondcluster}) and as a function of depth. 

The latter is fixed relative to the  characteristic  luminosity of the Schechter  function, $L^{\star}$.   $L^{\star}$, or  in this  case $m^{\star}$, is estimated by fitting a Schechter function to the statistically background corrected cluster luminosity function (as in \citealt{2011ApJ...734....3Z}). The limiting depth for the fit is chosen at $m^{\star}+4$, where $m^{\star}$ is found iteratively. The offset of 4 limits the magnitude range so it is well sampled for both clusters. In this way we find characteristic magnitudes of $m^{\star}_{R,\mathrm{Vega}}=19.3^{+0.5}_{-0.6}$ ($m^{\star}_{R,\mathrm{Vega}}=22.2^{+0.9}_{-1.7}$ for the $z\approx0.6$ cluster) for apparent \textit{R} band and $M^{\star}_{r,\mathrm{AB}}=-23.4^{+0.3}_{-0.3}$ ($M^{\star}_{r,\mathrm{AB}}=-23.9^{+0.9}_{-1.0}$) for absolute magnitudes in Sloan Digital Sky Survey (SDSS) \textit{r'} band.

We conclude that approximately $50$ per cent of cluster members are red galaxies near the core of RXC J2248.7-4431, and that this fraction continuously drops towards the outskirts, where it is below $20$ per cent. For the higher redshift system, the fraction of red galaxies in the cluster is significantly lower (which is known as the Butcher-Oemler effect, \citealt{1978ApJ...219...18B}). The fraction of red galaxies decreases both towards fainter magnitudes, larger separations from the core and higher redshift.

\subsection{Mass from richness and luminosity}
\label{sec:richness}
We can find an independent estimate of cluster mass by means of mass-observable relations of richness and absolute luminosity. To this end, we apply the prescription of \citet*{2010MNRAS.407..533W}. For our cluster member catalogue we select luminous galaxies inside a projected radius of 1Mpc around the BCG within $|z-z_{\rm cl}|\leq0.06(1+z_{\mathrm{cl}})$ and with magnitudes $M_{r'}\leq-21$, where we use absolute magnitudes in SDSS \textit{r'} band \citep{1998AJ....116.3040G}. The size and \textit{r'} band luminosity of the sample is 64 and $L_{r'}=3.0\times10^{12} \mathrm{L}_{\odot}$. After subtracting the mean number and luminosity of similar object selected from a region of the field more than 3~Mpc from the BCG, we find a background corrected richness of $R=54$ and luminosity of $L_{r'}=2.6\times10^{12} \mathrm{L}_{\odot}$.

Applying the scaling relations of \citet{2010MNRAS.407..533W}, this yields
\begin{eqnarray}
\log M_{101c}/(10^{14} h^{-1} \Msol) &=& (-1.57\pm0.12)+1.55\log R  \nonumber \\ 
&=& 1.12\pm_{\rm stat}0.09\pm_{\rm sys}0.12
\end{eqnarray}
from richness as the mass proxy with systematic and statistical uncertainty. Alternatively, one can use the net luminosity to obtain
\begin{eqnarray}
\log M_{101c}/(10^{14} h^{-1} \Msol) &=& (-2.03\pm0.06) \nonumber \\ &+& 1.49\log L_{r'}/(10^{10}h^{-2}\mathrm{L}_{\odot}) \nonumber \\ &=& 1.14\pm_{\rm stat}0.09\pm_{\rm sys}0.06 \; .
\label{eqn:mlm}
\end{eqnarray}
Since the errors of these two estimates are highly correlated we use the less uncertain equation~(\ref{eqn:mlm}), which yields a mass of $M_{101c}=19\pm6\times10^{14}\Msol$.

\section{Weak Lensing Measurement}

Weak gravitational lensing changes the ellipticities of the images of galaxies in the background of massive structures. Its measurement therefore requires the determination of pre-seeing galaxy shapes. In this section we describe our shape measurement, while the analysis of the signal is presented in Section~\ref{sec:wlanalysis}.

\subsection{Model of the point spread function}

\begin{figure}
\centering
\includegraphics[width=0.48\textwidth]{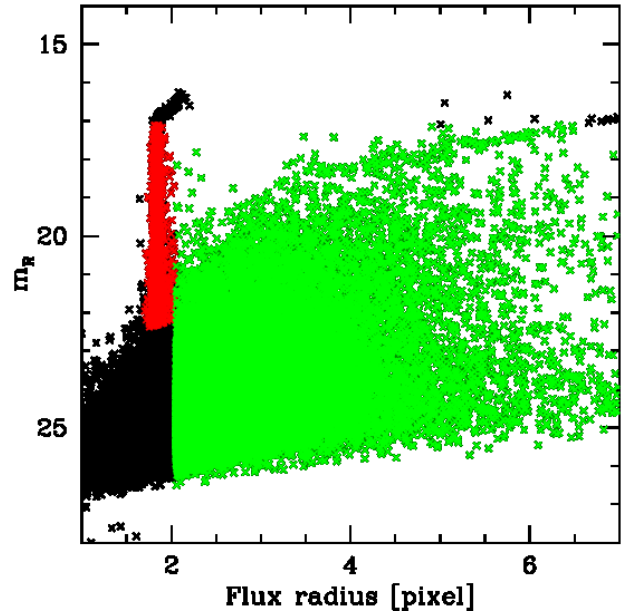}
\caption{Selected stars for PSF modelling (red) and sources initially accepted for shape measurement (green).}
\label{fig:starselect}
\end{figure}
\begin{figure}
\centering
\includegraphics[width=0.48\textwidth]{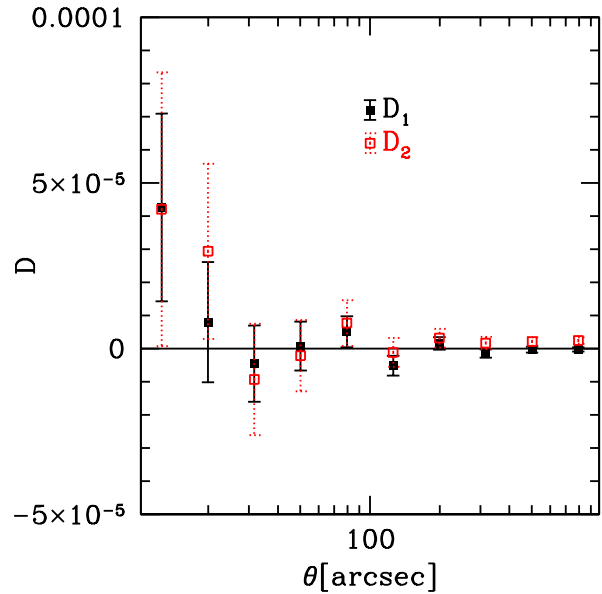}
\caption{Autocorrelation of PSF model ellipticity residuals ($D_1$) and cross-correlation of residuals and measured ellipticity ($D_2$) according to \citet{2010MNRAS.404..350R} for a fifth order polynomial PSF model in the \textit{R} band co-add. While lower orders show signs of underfitting ($D_1$ and/or $D_2$ significantly non-zero on some scales), higher orders do not improve the model further and show signs of overfitting ($D_2$ significantly negative on some scales).}
\label{fig:psfdiag}
\end{figure}

Any estimate of the pre-seeing shape of a galaxy requires knowledge of the point spread function (PSF). For a successful and unbiased weak lensing analysis it is most crucial to model the PSF accurately \citep[cf.][]{2013ApJS..205...12K}, since mismatches in ellipticity and size of the PSF model cause additive and multiplicative systematic errors in the shape estimate, respectively.

We use PSFEx\footnote{\texttt{http://www.astromatic.net/software/psfex}} \citep{2011ASPC..442..435B} for determining the profile of the PSF as a polynomial function of the position in the co-added focal plane. To this end, we perform a pre-selection of stars according to size, S/N and the \textsc{SExtractor}\footnote{\texttt{http://www.astromatic.net/software/sextractor}} neural network star classifier \citep{1996A&AS..117..393B}. The pre-selected stars (cf. Fig.~\ref{fig:starselect}) are used to determine the PSF model, which is then checked using the following diagnostics:
\begin{enumerate}[i]
 \item mean and scatter of residual ellipticities and sizes to check for remaining offsets and quality of the fit;
 \item whisker plot of residual ellipticities to exclude failure of the fit in specific regions of the focal plane;
 \item two-point autocorrelations of ellipticity residuals ($D_1$) and cross-correlation of residuals and measured star ellipticities ($D_2$) to exclude under- and overfitting, as defined in \citet{2010MNRAS.404..350R}, equations (13) and (14).
\end{enumerate}

We verify, in particular, that a star pre-selection is necessary for the size of the PSFEx model to match the size of the PSF well enough. In \textit{R} and \textit{V} a smooth PSF model over the whole focal plane is a sufficient description of the observed pattern. We find that, similar to other exposures from the WFI camera we have analysed in the past, the \textit{I}-band PSF is more difficult to model. Only when masking the border regions between neighbouring chips and discarding all stars and galaxies in those regions can we describe the PSF with a simple polynomial dependence on position to sufficient accuracy. For these tests we find that the method of \citet{2010MNRAS.404..350R} is particularly helpful, and use it to determine the correct polynomial order of the spatial variation [5 (7) for the \textit{R} (\textit{V} and \textit{I}) band, respectively], which match the observed PSF without clear signs of over- or underfitting (cf. Fig.~\ref{fig:psfdiag}).

\subsection{Shape Measurement}

We run an implementation of the KSB+ method (\citealt*{1995ApJ...449..460K}, hereafter KSB; \citealt{1997ApJ...475...20L,1998ApJ...504..636H}) using the PSFEx PSF model (KSBPSFEx) that has been tested against simulations at intermediate to large tangential shears and proven to be viable in the cluster shear regime (Young et al., in preparation). The pipeline includes the following preparation steps for the KSB+ shape measurement:
\begin{enumerate}[i]
 \item unsaturated sources with flux radii larger than the stellar flux radius and zero \textsc{SExtractor} flags are filtered;
 \item postage stamps of 64 $\times$ 64~pixels size are extracted and neighbouring objects according to the \textsc{SExtractor} segmentation map are masked;
 \item the \textsc{SExtractor} photometric background estimate at the object position is subtracted from the image in order to compensate for small-scale background variations insufficiently modelled by the data reduction pipeline;
 \item bad and masked pixels are interpolated using a Gauss-Laguerre model of the galaxy \citep{2002AJ....123..583B}, where we discard objects with more than 20 per cent of postage stamp area or 5 per cent of model flux falling on to bad or masked pixels.
\end{enumerate}
As the final step of the shape measurement, KSB+ is run on the cleaned postage stamp of the galaxy and the sub-pixel resolution PSF model at the galaxy position. For details, we refer the reader to the papers introducing and extending the method. Here we only give a brief summary.

KSB measure polarizations,
\begin{equation}
\bm{e} = \frac{1}{Q_{11}+Q_{22}} \left(
\begin{array}{c}
Q_{11}-Q_{22}\\
2Q_{12}
\end{array} \right) \; ,
\end{equation}
using second moments calculated inside a Gaussian aperture,
\begin{equation}
Q_{ij}=\int d^2\theta I(\bm{\theta}) w(|\theta|) \theta_i \theta_j \; ,
\end{equation}
of the surface brightness distribution of the galaxy $I(\bm{\theta})$ with a Gaussian weight function $w(|\theta|)$ centred on the galaxy centroid. In our implementation, the weight function for measuring the galaxy and PSF moments is scaled with the measured half-light radius of the observed galaxy.

Cluster weak lensing analyses require the ensemble measurement of reduced shear \citep[cf.][p.60]{2001PhR...340..291B}. In the presence of an elliptical PSF, the linear approximation of how observed post-seeing polarization $\bm{e_o}$ reacts to a reduced shear $\bm{g}$ can be expressed as
\begin{equation}
 \bm{e_o}=\bm{e}_i + \bm{\mathrm{P}}^{\mathrm{sm}}\bm{p}+\bm{\mathrm{P}}^{\gamma}\bm{g} \; ,
\end{equation}
where $\bm{e}_i$ is the intrinsic post-seeing ellipticity of the galaxy, $\bm{\mathrm{P}}^{\mathrm{sm}}$ is a $2\times2$ tensor quantifying the response of observed shear to PSF polarisation $\bm{p}$ and $\bm{\mathrm{P}}^{\gamma}$ is the shear responsivity tensor. Inverting $(\bm{\mathrm{P}}^{\gamma})^{-1}\approx\frac{2}{\tr \bm{\mathrm{P}}^{\gamma}}$ and assuming that $(\bm{\mathrm{P}}^{\gamma})^{-1}\bm{e}_i$ is zero on average because of the random intrinsic orientation of galaxies, this yields the ensemble shear estimate
\begin{equation}
 \langle\bm{g}\rangle=\langle\bm{\epsilon}\rangle=\left\langle \frac{2}{\tr \bm{\mathrm{P}}^{\gamma}}\left( \bm{e_o}-\bm{\mathrm{P}}^{\mathrm{sm}}\bm{p} \right)\right\rangle \; .
\label{eqn:gfrome}
\end{equation}
We remove objects with failed KSB+ measurements and objects with $\tr \bm{\mathrm{P}}^{\gamma}<0.1$ from the final shape catalogue. The latter is a requirement owing to the noisiness of measured $\tr \bm{\mathrm{P}}^{\gamma}$, which otherwise greatly amplifies the uncertainty of equation~(\ref{eqn:gfrome}) \citep[cf.][for similar clipping schemes at various levels of $\tr \bm{\mathrm{P}}^{\gamma}$]{2007AuA...468..823S,2010AuA...514A..88R,2012arXiv1208.0605A,2012arXiv1209.1391N}.  We match \textit{R}, \textit{V} and \textit{I} band shape catalogues and take the arithmetic mean of all available bands as the individual object shape and match against the photometric redshift catalogue to select valid background galaxies and assign redshift scalings. This and the S/N cut introduced in Section~\ref{sec:noisecal} leaves us with an average of nine galaxies per square arcminute in the background of RXC J2248.7-4431 with a mean ratio of angular diameter distances of $\langle D_{ds}/D_{s}\rangle=0.59$ (cf. Section~\ref{sec:wlintro}).

\subsection{Noise bias calibration}
\label{sec:noisecal}

Biases of shape estimators are commonly expressed as a multiplicative and an additive term \citep{2006MNRAS.368.1323H}, where the additive component is typically due to imperfect correction for the ellipticity of the PSF. The latter effect is particularly relevant for cosmic shear analyses, where auto-correlations of the shear field are calculated and auto-correlations of the PSF ellipticity field would enter the equation by means of an additive bias \citep[for a more detailed study of the influence of shape biases on cosmic shear, see][]{2008MNRAS.391..228A}. All analyses we do in this cluster weak lensing study, however, deal with shear estimates averaged in an annulus, in which case any constant additive bias cancels out. Pixel noise, on the contrary, typically causes multiplicative biases at the per cent level over a wide range of observational conditions, which would directly enter our cluster weak lensing model \citep{2012MNRAS.424.2757M}. This and the fact that we do not observe significant additive biases on simulated images with elliptical PSF is the reason for limiting our calibration to the most important factor, namely noise-dependent multiplicative bias.

\begin{figure}
\centering
\includegraphics[width=0.48\textwidth]{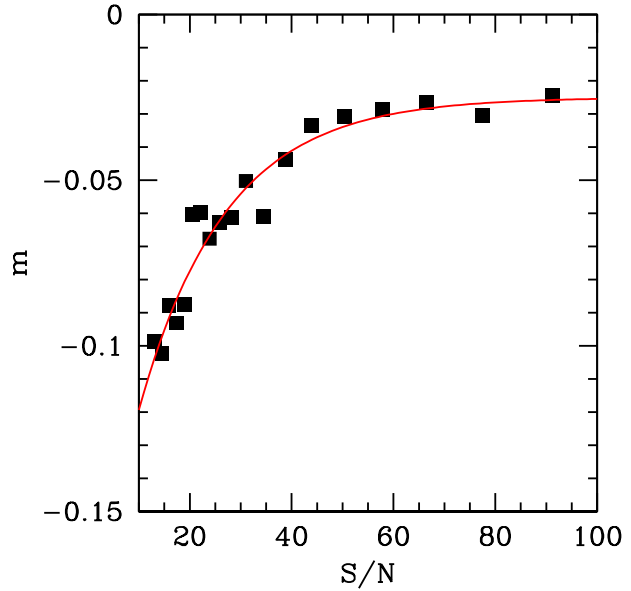}
\caption{Multiplicative bias $m$ as a function of S/N for our shape pipeline, as measured using image simulations with a range of PSF sizes, shears and galaxy properties typical for a weak lensing cluster study like ours. The results are well fit by equation~(\ref{eqn:sncal}) with best-fitting parameters $A=0.025$, $B=0.17$ and $C=17$ (red line).}
\label{fig:msn}
\end{figure}

Shape measurement, primarily because it involves division by noisy quantities, suffers from a noise bias, which is negligible for bright galaxies but becomes troublesome for the faint end of background galaxies in a weak lensing analysis \citep{2000ApJ...537..555K,2002AJ....123..583B,2012MNRAS.424.2757M,2012MNRAS.425.1951R,2012arXiv1203.5049K}. Many studies in the past have applied a global correction factor to their KSB shape catalogue (see \citealt{2007AuA...468..823S,2012ApJ...758..128C,2012ApJ...744...94R,2012arXiv1209.1391N} for a range of factors used), but there also have been cases where calibration has been matched to galaxy S/N and sometimes also size using simulated galaxy images \citep{2010AuA...516A..63S,2012arXiv1208.0597V}. We quantify the magnitude of noise bias in our pipeline using image simulations with a range of PSF sizes, shears and galaxy properties typical for a weak lensing cluster study like ours (Young et al., in preparation).

If a shear estimator $g_{o}$ only has multiplicative bias $m$, it can be written as
\begin{equation}
g_{o}=(1+m) g + N \; ,
\end{equation}
where $g$ is the true shear and $N$ is a noise term with zero mean. We fit $m$ for simulated sets of galaxies and show the results in Fig.~\ref{fig:msn} as a function of S/N measured according to the prescription of \citet{2001AuA...366..717E}. The multiplicative bias is fit well by the functional form
\begin{equation}
m = -A - B \exp(-(\mathrm{S/N})/C) \; ,
\label{eqn:sncal}
\end{equation}
with best-fitting values of $A=0.025$, $B=0.17$ and $C=17$ down to a S/N of 10.

We discard objects outside the regime this calibration was tested on (i.e. where S/N$<10$ \citep{2001AuA...366..717E} or (S/N)$_{\rm iso}<15$ as measured inside an isophotal aperture of $1.5\sigma$ significance per pixel over the background). The remainder of our sample is calibrated with the $m$ of equation~(\ref{eqn:sncal}), by multiplying shape estimates by a factor of $1/(1+m(\mathrm{S/N}))$.

\section{Weak Lensing Analysis}
\label{sec:wlanalysis}

\subsection{Introduction}
\label{sec:wlintro}

Cluster weak lensing analysis aims to reconstruct properties of the density field of clusters of galaxies from the reduced shear they impose on the images of background galaxies. We refer the reader to the review of \citet{2001PhR...340..291B} for an in-depth introduction and only give a brief overview of the main concepts here.

Gravitational shear $\bm{\gamma}$ relates to reduced shear $\bm{g}$ as
\begin{equation}
 \bm{\gamma}=\bm{g}(1-\kappa) \; ,
\label{eqn:reducedshear}
\end{equation}
where $\kappa$ is the projected surface mass density in units of the critical surface density, $\kappa=\Sigma/\Sigma_c$, with
\begin{equation}
 \Sigma_c=\frac{c^2}{4\pi G} \frac{D_s}{D_d D_{ds}} \; .
\end{equation}
The latter contains a geometric factor composed of angular diameter distances $D_{s,d,ds}$ from observer to source, from observer to lens and from lens to source, respectively.

The surface mass density is related to the mean tangential component of the shear on a circle, $\gamma_t(r)$, by the simple equation
\begin{equation}
 \gamma_t(r)=\langle\kappa(<r)\rangle-\kappa(r) \; ,
\label{eqn:gk}
\end{equation}
the difference between mean $\kappa$ inside and on the edge of the circle.

Equations~(\ref{eqn:reducedshear}) and (\ref{eqn:gk}) show that the observable $\bm{g}$ is invariant under mass sheet transformations, $\kappa\rightarrow\lambda\kappa+(1-\lambda)$, a degeneracy which can be broken for instance by assuming a functional form of $\kappa(r)$.

In the following we use two common density profiles, the singular isothermal sphere (SIS) and the NFW profile \citep*{1996ApJ...462..563N}. Their densities are given by
\begin{equation}
 \rho_{\mathrm{SIS}}(r)=\frac{\sigma_v^2}{2\pi G r^2} \; ,
\label{eqn:sisdens}
\end{equation}
with a constant velocity dispersion $\sigma_v$, and
\begin{equation}
 \rho_{\mathrm{NFW}}(r)=\frac{\rho_0}{(r/r_s)(1+r/r_s)^2} \; .
\end{equation}
The NFW profile has two free parameters which are commonly expressed in terms of the mass and concentration $c_{\Delta m}=r_{\Delta m}/r_s$ or $c_{\Delta c}=r_{\Delta c}/r_s$ instead of the central density $\rho_0$ and the scale radius $r_s$ used in the equation above.

The projected surface mass density and tangential shear profiles of a SIS can be readily integrated as
\begin{equation}
\kappa(r)=\gamma_t(r)=2\pi\left(\frac{\sigma_v}{c}\right)^2 \frac{D_{ds}}{D_s}\frac{1}{r} \; .
\label{eqn:sis}
\end{equation}
For the NFW profile, we refer the reader to the calculations in \citet{2000ApJ...534...34W}.

\subsection{Tangential alignment}

\begin{figure}
\centering
\includegraphics[width=0.48\textwidth]{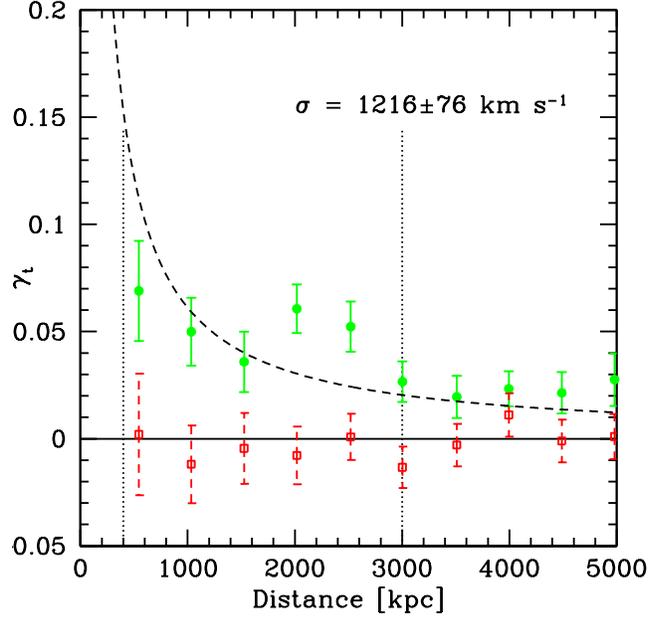}
\caption{Tangential alignment (green circles), \textit{B} mode (red squares) and SIS fit (dashed line) inside the interval delimited by dotted vertical lines. Note that the \textit{B} mode is consistent with 0 as expected for a bias-free shape catalogue.}
\label{fig:sis}
\end{figure}

We first measure the tangential alignment signal around the brightest cluster galaxy (cf. Fig.~\ref{fig:sis}). For this we calculate the mean reduced tangential shear $g_t$ in radial bins and convert it into tangential gravitational shear $\gamma_t$ assuming a SIS profile, for which (cf. equations~\ref{eqn:reducedshear} and \ref{eqn:sis})
\begin{equation}
\gamma_t = \frac{g_t}{1+g_t} \; .
\end{equation}
The effective distance ratio $D_{ds}/D_{s}$ is calculated from the mean individual galaxy values in each bin. The SIS fit yields a velocity dispersion of $\sigma_v=1216\pm76\ \mathrm{km}\ \mathrm{s}^{-1}$, measuring in the radial range of 400-3000 kpc projected distance. As it follows from equation~(\ref{eqn:sisdens}),
\begin{eqnarray}
r_{200c} &=& \frac{\sqrt{2}\sigma_v}{10H(z)} = (2025\pm126)\;\mathrm{kpc}\;\; \mathrm{and}\\
r_{200m} &=& \frac{\sqrt{2}\sigma_v}{10H_0\sqrt{\Omega_m(1+z)^3}} = (2979\pm184)\;\mathrm{kpc} \; ,
\end{eqnarray}
which implies a mass of $M_{200c,\rm SIS}=(14.0\pm2.6) \times 10^{14}\ M_{\odot}$ and $M_{200m,\rm SIS}=(21.0\pm3.8) \times 10^{14}\ M_{\odot}$.

\subsection{Significance map}

\begin{figure}
\centering
\includegraphics[width=0.5\textwidth]{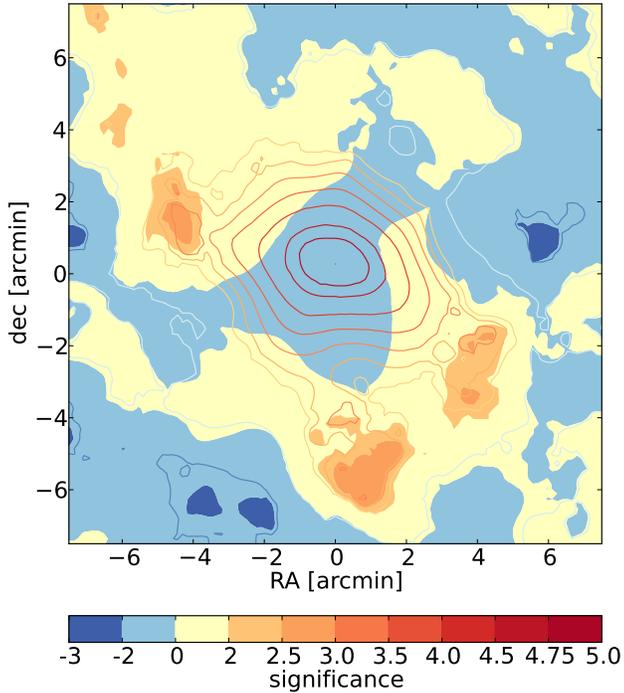}
\caption{Aperture mass significance map calculated according to equations~(\ref{eqn:wmap}) and (\ref{eqn:map}). Contour lines show the significance of aperture mass measured on background galaxies according to the cut of equation~(\ref{eqn:zcut}), which is centred on the BCG at a significance of $5\sigma$. The background colour shows the aperture mass significance of the residual shape catalogue after subtracting the best-fitting NFW model of the central halo (cf. Section~\ref{sec:nfw}). Towards the lower left is the south-eastern corner of the image.}
\label{fig:map}
\end{figure}

For a first two-dimensional view of the lensing signal, we measure the surface mass density or, equivalently, tangential gravitational shear inside circular weighted apertures, so-called aperture masses \citep{1996MNRAS.283..837S}. We show the significance of aperture masses above zero as a function of position in Fig.~\ref{fig:map}. For this we use a Gaussian weight function
\begin{equation}
 w(|\bm{\theta}|)\propto\left\{
  \begin{array}{ll}
    \exp(-|\bm{\theta}|^2/(2\sigma_w^2)) & |\bm{\theta}|<3\sigma_w \; , \\
    0 & \mathrm{otherwise}
  \end{array}
\right.
\label{eqn:wmap}
\end{equation}
to calculate the significance, defined as the ratio between aperture mass and its uncertainty, $M_{\rm ap}/\sigma_{M_{\rm ap}}$ (\citealt{2001PhR...340..291B}, their Section~5.3 and  \citealt{2004AuA...420...75S}), with
\begin{eqnarray}
 M_{\rm ap}(\bm{\theta})&=&\sum_i w(|\bm{\theta}-\bm{\theta_i}|) g_{i,t} \nonumber \\
 \sigma_{M_{\rm ap}}&=&\sqrt{\frac{1}{2}\sum_i w^2(|\bm{\theta}-\bm{\theta_i}|) |g_i|^2} \; ,
 \label{eqn:map}
\end{eqnarray}
where $g_{i,t}$ is the tangential reduced shear of galaxy $i$ measured with respect to $\bm{\theta}$. For the width of the aperture we use $\sigma_w=3$~arcmin.
We detect the cluster as a peak in the aperture significance map with a significance of 5$\sigma$, centred on the BCG with a deviation of only few arcseconds. The significance map indicates an anisotropic distribution of mass around its centre, with additional peaks towards the north-eastern, south-western and southern direction from the BCG. We also check the significance of aperture \textit{B} mode peaks and find them to be consistent with a random field.

\subsection{Mass density map}
\label{sec:kappamap}
\begin{figure}
\centering
\includegraphics[width=0.48\textwidth]{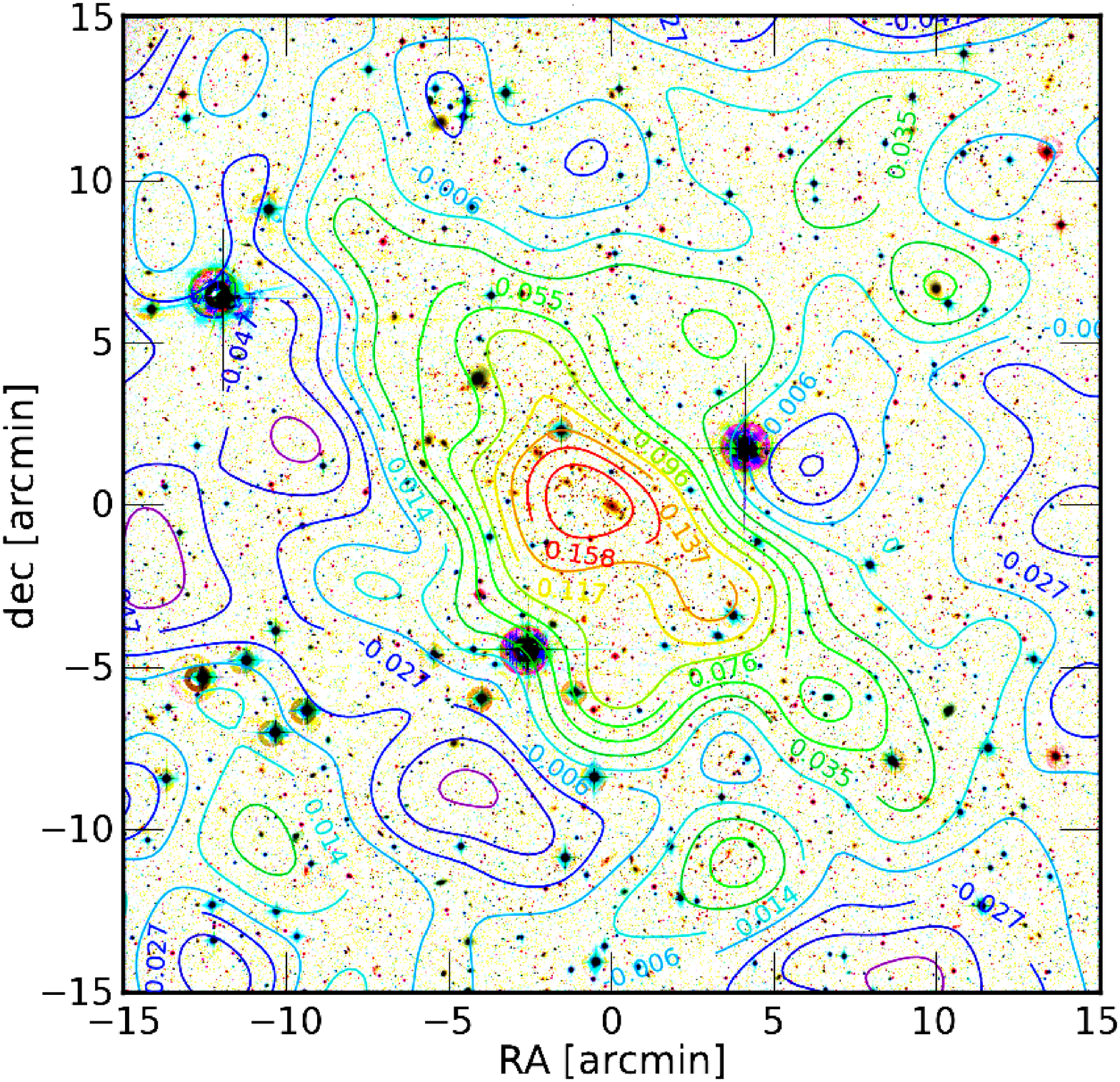}
\caption{Density map, overlayed on colour image of RXC J2248.7-4431.}
\label{fig:massmap}
\end{figure}

We calculate a density map using the method of \citet{1993ApJ...404..441K}. For this we use all sources that satisfy the background cut of equation~(\ref{eqn:zcut}). These sources have a mean ratio of angular diameter distances of $\langle D_{ds}/D_{s}\rangle=0.59$. The $\kappa$ map has a pixel size of 1~arcmin and is smoothed with a Gaussian of $\sigma=1.2$~arcmin width. The map is shown in Fig.~\ref{fig:massmap} for the full field of view. 

\subsection{NFW model}
\label{sec:nfw}

\begin{table*}
\begin{center}
\begin{tabular}{|l|r|r|l|l|}
\hline
Mass definition & Our NFW fit & Literature value & Source & Method \\ \hline
$M_{200m}$ & $33.1^{+9.6}_{-6.8}$ & $28.2\pm_{\rm stat}3.6\pm_{\rm sys}9.3$ & \citet{2011ApJ...738..139W} & SZ \\ \hline
$M_{101c}$ & $32.2^{+9.3}_{-6.6}$ & $19\pm6$ & This work, Section~\ref{sec:richness}, and \citet{2010MNRAS.407..533W} & Luminosity \\ \hline
$M_{200c}$ & $22.8^{+6.6}_{-4.7}$ & $42^{+17}_{-9}$ & \citet{2012AJ....144...79G} & Kinematics \\ \hline
$M_{500c}$ & $12.7^{+3.7}_{-2.6}$ & $12.25\pm0.21$ & \citet{2011AuA...536A..11P} & X-ray / \textit{XMM-Newton} \\ 
& & $11.5\pm_{\rm stat}2.6\pm_{\rm sys}0.5$ & \citet{2011AuA...536A..11P} & SZ \\ \hline
$M_{2500c}$ & $2.8^{+0.8}_{-0.6}$ & $5.3\pm2.6$ & \citet{2011MNRAS.418.1089C} & X-ray / \textit{Chandra} \\ \hline
\end{tabular}
\end{center}
\caption{Best-fitting mass and confidence region of our NFW model ($M_{200m}$) converted to other definitions of mass inside spherical overdensity using the best-fitting concentration of $c_{200m}=2.6$. All values are given in units of $10^{14}\Msol$.}
\label{tbl:mass}
\end{table*}

\begin{figure}
\centering
\includegraphics[width=0.48\textwidth]{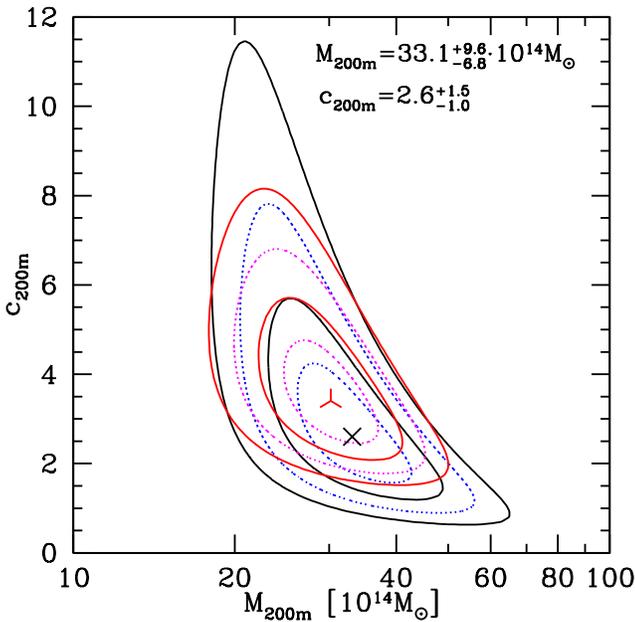}
\caption{Likelihood contours of the two-parametric NFW profile of RXC J2248.7-4431. Black, solid (blue, dotted) contours show the combined (projected) confidence regions (intervals) for $M_{200m}$ and $c_{200m}$ with no concentration prior. Red, solid (magenta, dotted) contours for the combined (projected) confidence regions (intervals) using the concentration prior of \citet{2001MNRAS.321..559B} and \citet{2008MNRAS.390L..64D}. All contours are drawn at the $1\sigma$ and $2\sigma$ confidence levels. The black cross (red triangular symbol) indicate the best-fitting solution without (with) concentration prior.}
\label{fig:nfw}
\end{figure}
\begin{figure}
\centering
\includegraphics[width=0.48\textwidth]{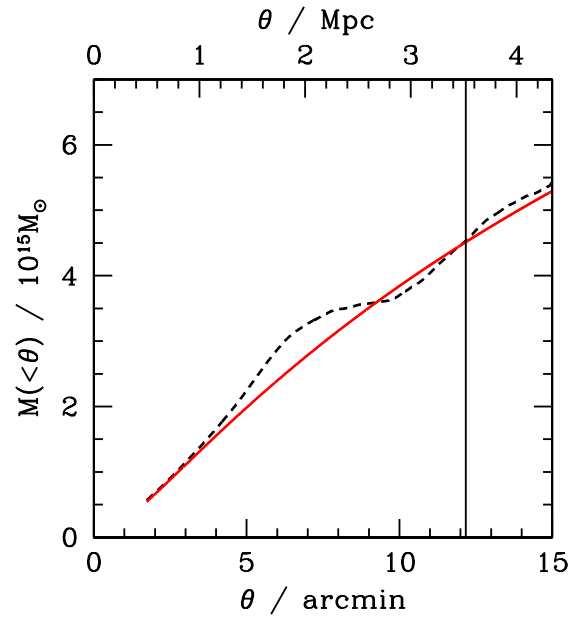}
\caption{Projected excess mass inside a circle around the BCG of RXC J2248.7-4431, shown for the best-fitting NFW model (red, solid line) and directly measured from the observed tangential shear profile (black, dashed line), fixed at $r_{200m}$ (vertical line).}
\label{fig:minr}
\end{figure}

We further perform a likelihood analysis (see \citealt*{schneider00}) for a two-parametric NFW profile \citep{1996ApJ...462..563N} of the halo, fitting the concentration parameter $c_{200m}$ and the mass $M_{200m}$ simultaneously.  We use background galaxies above a projected separation of 1~Mpc at the cluster redshift, since inside this limit some of the tested models predict shear above the weak regime and, in addition, astrophysical effects near the core of the cluster disturb the expected signal most \citep[cf.][]{2010MNRAS.405.2078M}. This justifies the assumption of a model-independent dispersion of intrinsic shapes in equation~(15) of \citet{schneider00}. We use $\sigma_{\epsilon}=0.3$, the dispersion of our best-fitting subtracted shape residuals, and determine confidence limits using the $\Delta\chi^2$ statistics of \citet{1976ApJ...210..642A}.

The $1\sigma$ confidence limits for mass and concentration individually are $M_{200m}=33.1^{+9.6}_{-6.8}\times 10^{14}\Msol$, $c_{200m}=2.6^{+1.5}_{-1.0}$, where because of a degeneracy the very high masses only occur at unlikely low concentrations. Confidence contours are shown in Fig.~\ref{fig:nfw}.

Since there is prior knowledge about the concentration of dark matter haloes at given mass and redshift, we can reduce the uncertainty of our measurement by multiplying the likelihood with a concentration term
\begin{equation}
p(c|M,z)=e^{-(\log c-\log c(M,z))^2/(2\sigma_{\log c}^2)} \; ,
\label{eqn:cm}
\end{equation}
where we adopt the mass-concentration relation $c(M,z)$ of \citet{2008MNRAS.390L..64D} and a lognormal distribution of concentrations according to \citet{2001MNRAS.321..559B} with $\sigma_{\log c}=0.18$. This leads to confidence limits of $M_{200m}=30.2^{+6.1}_{-5.1}\times10^{14}\Msol$ and $c_{200m}=3.4^{+1.3}_{-0.9}$.

For reference and comparison, we give corresponding masses in different definitions of spherical overdensity in Table~\ref{tbl:mass}. Our result is consistent inside the error limits with the X-ray mass of the \citet{2011AuA...536A..11P} and the SZ masses of \citet{2011ApJ...738..139W} and \citet{2011AuA...536A..11P}. The X-ray mass at high overdensity of \citet{2011MNRAS.418.1089C} is higher than our best fit, yet relatively uncertain and likely dominated by highly concentrated baryonic matter in and around the BCG, which is not correctly modelled by a global NFW halo. The mass calculated from richness and luminosity in Section~\ref{sec:richness} is significantly lower than our weak lensing result, yet likely contains systematic uncertainties because it was calibrated with different data. Spectroscopic velocity dispersion of cluster members yields a significantly higher mass estimate \citep{2012AJ....144...79G}, discussed in more detail in Section~\ref{sec:gomez}. In addition, we plot mass enclosed in a cylinder of varying radius, centred on the BCG, in Fig.~\ref{fig:minr}.

\begin{figure*}
\centering
\includegraphics[width=0.48\textwidth]{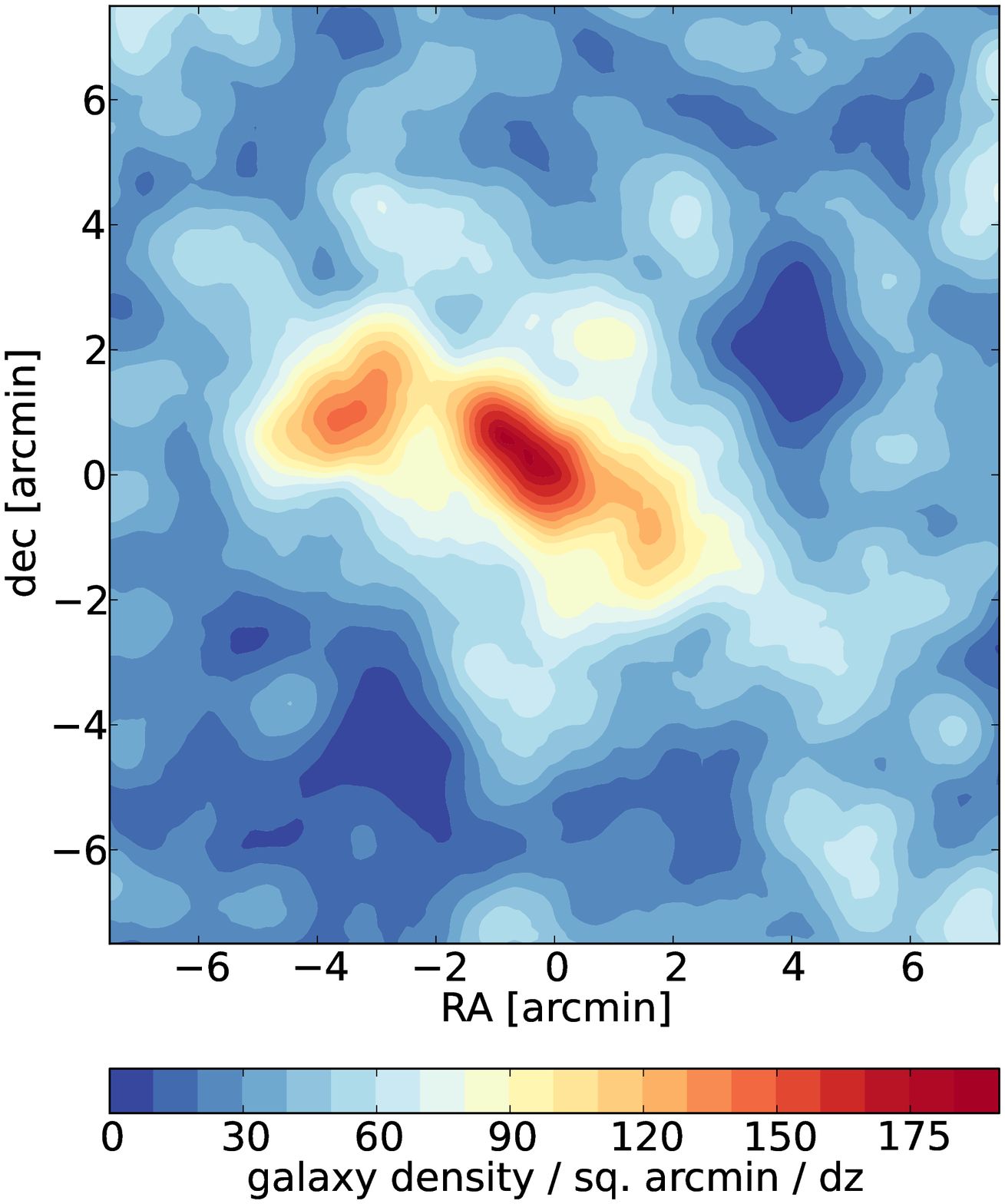}
\includegraphics[width=0.48\textwidth]{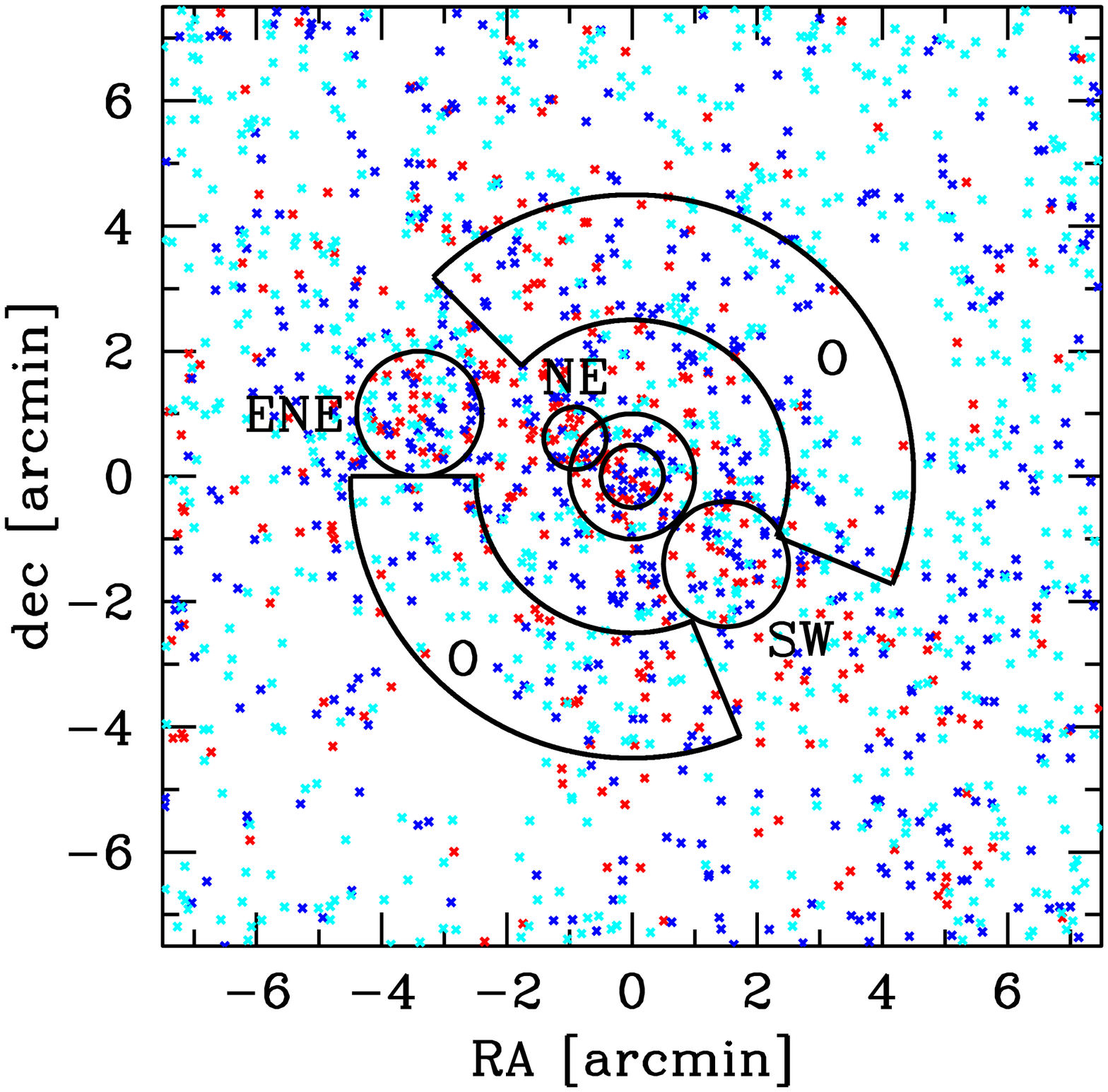}
\caption{Left-hand panel: density of galaxies with $|z-z_{\mathrm{cl}}|<0.06(1+z_{\mathrm{cl}})$ in our photometric redshift catalogue centred on the BCG position, measured as a kernel density with the minimum-variance (Epanechnikov) kernel of 1~arcmin radius. Right-hand panel: positions of red (red), moderately star-forming Sb/c-like (blue) and star-forming (cyan) galaxies around the centre of RXC J2248.7-4431 in the same redshift slice. The areas delineated by black lines indicate the regions used for the SED analysis in Table~\ref{tbl:seds}. The two empty regions visible at radii of $\approx$5~arcmin from the center correspond to positions of bright stars and their associated masks.}
\label{fig:possed}
\end{figure*}

Deviations from the spherically symmetrical NFW profile are known to exist in several different forms, such as correlated secondary haloes, asphericity of the main halo or filamentary structures \citep{2012Natur.487..202D} and can influence the accuracy of weak lensing cluster analyses significantly \citep{2011MNRAS.416.1392G}. In order to investigate these effects, we plot the aperture mass significance map on a shape catalogue with the signal of the best-fitting model subtracted in Fig.~\ref{fig:map}. While the central peak is described well by the spherical profile, we find that there are additional peaks towards the north-eastern, south-western and southern direction from the BCG, with the northern one potentially highly elongated towards the north-north-east. This is in line with the density map (cf. Section~\ref{sec:kappamap}) and the shape of the galaxy concentration in three dimensional space (cf. Fig.~\ref{fig:zdens}). These structures therefore most likely represent correlated structures physically close to the main halo of RXC J2248.7-4431.

\section{On the hypothesis of a recent merger in RXC J2248.7-4431}
\label{sec:gomez}

\citet{2012AJ....144...79G} present evidence for a recent merger of RXC J2248.7-4431 based on optical imaging, spectroscopy of 51 cluster members and analysis of the X-ray emission of the system. We briefly recollect their findings and compare them with our results from deeper images with photometric redshift information and weak lensing.

\begin{table}
\begin{center}
\begin{tabular}{|l|r|r|r|r|}
\hline
Peak & $N_{\rm gal}$ & Red [\%] & Mod. SF [\%] & SF [\%] \\ \hline
Central 1' & 78 & $41\pm6$ & $41\pm6$ & $18\pm4$ \\ \hline
ENE & 67 & $33\pm6$ & $37\pm6$ & $30\pm6$ \\ \hline
SW & 50 & $32\pm7$ & $36\pm7$ & $32\pm7$ \\ \hline
O & 239 & $22\pm3$ & $36\pm3$ & $42\pm3$ \\ \hline
Central 0.5' & 29 & $48\pm9$ & $41\pm9$ & $11\pm6$ \\ \hline
NE & 29 & $51\pm9$ & $28\pm8$ & $21\pm8$ \\ \hline
\end{tabular}
\end{center}
\caption{Fractions of galaxies by SED type for the galaxy density peaks identified in Section~\ref{sec:numdens}.}
\label{tbl:seds}
\end{table}

\subsection{Galaxy number density}
\label{sec:numdens}

The cluster member number density, selected either based on spectroscopic redshift or in colour-magnitude space, is found by \citet{2012AJ....144...79G} to be multimodal (their Figs.~7-9). We study the projected density of galaxies in the region of the cluster in Fig.~\ref{fig:possed}. The multimodality is confirmed, where the two peaks of \citet{2012AJ....144...79G} form a highly enlongated central region of approximately 2~arcmin diameter. An additional peak towards the east-north-eastern (ENE) direction (not inside the field of view of the images used by \citealt{2012AJ....144...79G}) and a less dense but still visible peak towards the south-western (SW) direction (also identified by \citealt{2012AJ....144...79G}, but below their significance threshold). 

Our mean photometric redshifts of galaxies inside a cylinder of 1~arcmin radius around all three peaks are mutually consistent with each other and with the spectroscopic mean cluster redshift. The uncertainty of the ensemble mean of $\delta z\approx0.005$ for each peak, however, would only allow a significant detection of peculiar motions of several 1000~km~s$^{-1}$ along the line of sight.

Comparing the galaxy density map with Fig.~\ref{fig:map}, we find a rough correspondence of the two galaxy density sub-peaks with the two residual peaks in the aperture mass significance after subtracting the best-fitting model for the central halo. The third significant additional aperture significance peak towards the south does not correspond to a structure clearly visible in the galaxy number density.

\subsection{Galaxy SEDs}

We examine the distributions of best-fitting SED types from the photometric redshift code for the galaxy density peaks identified in \citet{2012AJ....144...79G} and in the previous section. To this end, we select galaxies inside cylindrical volumes of 1~arcmin (0.5~arcmin) radius and $|z-z_{cl}|<0.06(1+z_{cl})$ around the three peaks (and the two central peaks of \citealt{2012AJ....144...79G}). Results are shown in Fig.~\ref{fig:possed} and Table~\ref{tbl:seds}, discriminating red, moderately star-forming Sb/c-like and star-forming SEDs.

The galaxy populations in the ENE and SW peak contain a smaller fraction of red galaxies than in the very centre of the cluster, yet a larger one than elsewhere at similar separation from the BCG (in the region labelled O). 

Even if their distance from the centre in three dimensions is higher than in projection, the latter observation cannot be explained by the environmental influence of the core of RXC J2248.7-4431 alone and indicates that ENE and SW are evolved neighbouring structures in the outskirts of a major cluster.

For the two peaks identified by \citet{2012AJ....144...79G}, the galaxy populations are too small to make significant statements about differences in SED distribution (cf. Table~\ref{tbl:seds}, central 0.5~arcmin and NE).

\subsection{Centroid offsets}

\citet{2012AJ....144...79G} note an offset between the BCG and the centroid of X-ray emission when measuring the latter on larger radii. We note that the point of highest significance of the lensing signal (cf. Fig~\ref{fig:map}) is less than 20~arcsec off the BCG, which is consistent with the random offsets expected from shape noise \citep[cf.][]{2012MNRAS.419.3547D}. The strong lensing model of RXC J2248.7-4431 \citep{annamonna} also shows no offset between the central galaxy and the peak of the projected density field.

\subsection{Discrepancy between lensing and other mass estimates}
\label{sec:discrep}

Weak lensing mass estimates are not influenced by the astrophysical state of a system, unlike estimates based on the dynamic state of cluster members or the intra-cluster gas. Therefore a discrepancy between weak lensing and other mass estimates could suggest the influence of dynamic astrophysical processes such as mergers. \citet{2012AJ....144...79G} hypothesize that a lensing mass significantly lower than the dynamical and X-ray masses found for RXC J2248.7-4431 could help confirm merger activity.

The mass calculated by \citet{2012AJ....144...79G} based on the velocity dispersion of cluster members is slightly inconsistent with our weak lensing mass on the high side ($M^{\mathrm{dyn}}_{200c}=42^{+17}_{-9}\times10^{14}\Msun$ versus $M^{\mathrm{WL}}_{200c}=22.8^{+6.6}_{-4.7}\times10^{14}\Msol$), indicating that opposite bulk motion of different galaxy populations along the line of sight could be present. Whether there is indeed substructure in the velocity fields of the cluster members, for which \citet{2012AJ....144...79G} find marginal evidence will be investigated in detail with the forthcoming VLT-CLASH large spectroscopic programme (P. Rosati, private communication).

We see no significant discrepancy, however, of weak lensing with X-ray or SZ mass estimates of the cluster. \citet{2012AJ....144...79G} find an inconsistency between masses inside 110~kpc apertures from X-ray modelling in hydrostatic equilibrium and a strong lensing model. The latter, however, is based on a candidate multiply imaged system identified in their relatively shallow photometry which deeper observations have revealed to be not from the same source \citep{annamonna}. At this point, there is therefore also no compelling evidence of a discrepancy between lensing and X-ray mass estimates in the core of RXC J2248.7-4431.

\subsection{Influence of neighbouring structures on lensing mass}

We attempt to quantify the influence of the structures in the environment of RXC J2248.7-4431 on its weak lensing mass estimate by fitting a model with multiple haloes to the observed shear signal. To this end, we place two haloes with fixed concentration-mass relation \citep{2008MNRAS.390L..64D} at the position of the BCG and the centre of the east-northeastern peak (cf. Section~\ref{sec:numdens}). The confidence region for the mass of the secondary peak is in this case $M_{200m}^{ENE}=4.2^{+2.2}_{-1.9}\times 10^{14}\Msol$. Subtracting the best-fitting model of the secondary peak and fitting the central halo of RXC J2248.7-4431 with two free parameters yields $M_{200m}=24.0^{+6.1}_{-5.8}\times 10^{14}\Msol$ and $c_{200m}=2.8^{+3.0}_{-1.2}$. The corresponding mass for comparison with other estimates is $M_{500c}=9.6^{+2.5}_{-2.3}\times 10^{14}\Msol$.

We note that this lower mass of the central peak is still in agreement with the SZ and X-ray masses listed in Table~\ref{tbl:mass}. Is is unclear whether or not the mass of the secondary peak should in fact be included in a $M_{200m}$ estimate of RXC J2248.7-4431 since the projected separation of 3.5~arcmin is well below $r_{200m}$ of the central system. While this exemplifies the sensitivity of cluster weak lensing to correlated structures \citep{2011MNRAS.416.1392G}, it does not change the conclusions of Section~\ref{sec:discrep}.

\section{Secondary cluster at $z\approx0.6$}
\label{sec:secondcluster}
In redshift space density maps we detect a secondary peak around $(\alpha,\delta)=(22^{\mathrm{h}}49^{\mathrm{m}}37.1^{\mathrm{s}},-44^{\circ}43'04'')$, at the position of a bright early-type galaxy with photometric redshift of $z=0.66$ and corresponding absolute rest-frame magnitude of $M_{R,\mathrm{Vega}}=-25.1$ which appears to be at the centre of another cluster of galaxies. At a separation of 14.5~arcmin from the centre of RXC J2248.7-4431, this configuration of lenses is significantly more separated than the cases of cluster-cluster lensing discussed in \citet{2012MNRAS.420.1621Z}. In the following, we present a weak lensing and a simple strong lensing analysis of the newly discovered cluster.

\begin{figure}
\centering
\includegraphics[width=0.48\textwidth]{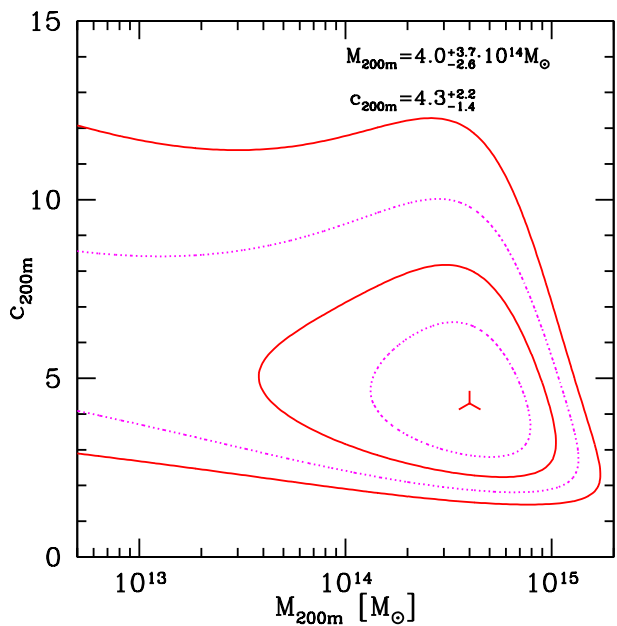}
\caption{Likelihood contours of the two-parametric NFW profile of secondary cluster with $z=0.58$. Red, solid (magenta, dotted) contours for the combined (projected) confidence regions (intervals) for $M_{200m}$ and $c_{200m}$ using the concentration prior of \citet{2001MNRAS.321..559B,2008MNRAS.390L..64D}. Both contours are drawn at the $1\sigma$ and $2\sigma$ confidence levels. The triangular symbol indicates the best-fitting solution.}
\label{fig:nfw2249}
\end{figure}

\begin{figure}
\centering
\includegraphics[width=0.48\textwidth]{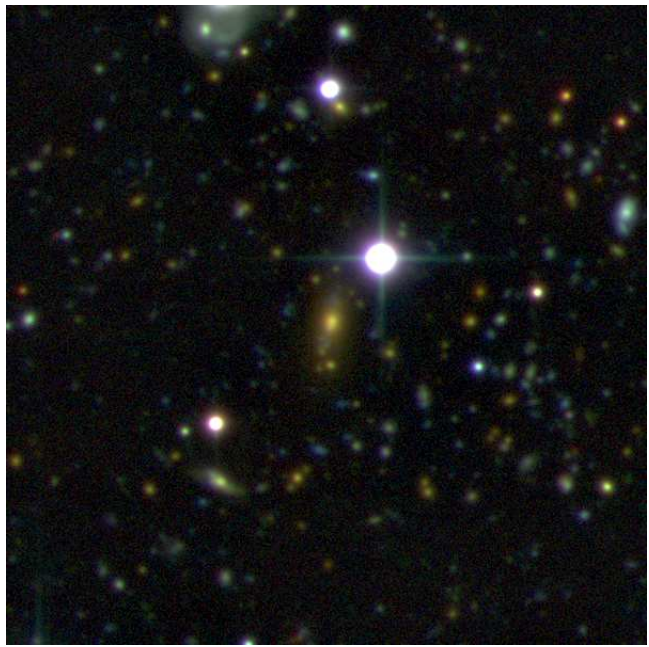}
\caption{Colour image of second cluster at $(\alpha,\delta)=(22^{\mathrm{h}}49^{\mathrm{m}}37.1^{\mathrm{s}}$, $-44^{\mathrm{\circ}}43'04'')$, $z\approx0.6$, with a size of 2$\times$2~arc\-min$^2$. Note the redder colour of BCG and cluster member galaxies compared to RXC J2248.7-4431 (cf. Fig.~\ref{fig:colorimage}) and the four symmetric blue sources along the major and minor axes of the BCG (the former ones highly blended with the BCG itself) at $z_{\mathrm{phot}}\approx1.5-1.8$.}
\label{fig:secondcluster}
\end{figure}

The mean redshift of the 12 closest, visually colour-selected member galaxies is $z=0.58$, which is the redshift we use for the following weak lensing analysis. After subtracting the best-fitting NFW model of the central peak of RXC J2248.7-4431 (cf. Section~\ref{sec:nfw}), we model this structure as an NFW halo positioned on the BCG using the concentration prior of equation~(\ref{eqn:cm}) and a radial range of $0.5\;\Mpc\leq r\leq3\;\Mpc$. The confidence contours are shown in Fig.~\ref{fig:nfw2249}. As a result of the low background number density and location of the cluster near the edge of the image, the mass is only weakly constrained and not detected at the $2\sigma$ level. We find the confidence region to be $M_{200m}=4.0^{+3.7}_{-2.6}\times10^{14}\Msol$ at a concentration of $c_{200m}=4.3^{+2.2}_{-1.4}$. Owing to the lack of completeness at the required depth, a richness and luminosity derived mass estimation (cf. Section~\ref{sec:richness}) is not feasible. Fig.~\ref{fig:secondcluster} shows a colour image of the central part of the system.

We identify four blue sources in a cross-like orientation around the BCG with separations of $\sim8$ and $\sim17$~arcsec along the small and large axis, respectively, as a candidate of multiple imaging. Assuming a cluster redshift of $z_l=0.58$, the redshift of the blue source at its photometric value of $z_s=1.58$, and a halo with projected ellipticity and orientation following the light distribution of the BCG, we can build a strong lensing model. For a non-singular halo with 4~arcsec (27 kpc) core radius, we get a mass of $\sim1.4\times10^{13}\Msol$ within a radius of 6.8~arcsec, the Einstein radius at this redshift.
                                                                                                                                                                                                                                                                                                                                                                                                                                                                                                                                                                                                                                                                                    

\section{Conclusions}

We present the first weak lensing analysis of the cluster of galaxies RXC J2248.7-4431. We calculate photometric redshifts of background and cluster galaxies and determine the richness and total luminosity of the system. We find the mass and concentration of RXC J2248.7-4431 in a NFW likelihood analysis to be equal to $M_{200m}=33.1^{+9.6}_{-6.8}\times10^{14}\Msol$, in good agreement with previous X-ray and SZ mass estimates, and $c_{200m}=2.6^{+1.5}_{-1.0}$, which is in the lower range of expected concentration for a system of the given mass and redshift. The subtraction of the best-fitting model leaves three marginally significant aperture mass peaks in the vicinity of the main halo, which likely correspond to substructure and surrounding structures of RXC J2248.7-4431. We confirm and add to some of the evidence given by \citet{2012AJ....144...79G} for an ongoing major merger of the system in terms of the multimodal galaxy density field and a discrepancy between (low) weak lensing and (high) dynamical mass estimate. We do not, however, find a tension between X-ray hydrostatic masses and the weak lensing mass of the cluster. This remains to be true even when a secondary peak of galaxy density at 3.5~arcmin separation of the core of RXC J2248.7-4431 is independently included in the weak lensing model.

We detect a second cluster at $z\approx0.6$ inside our field of view, whose weak lensing mass is weakly constrained at $M_{200m}=4.0^{+3.7}_{-2.6}\times10^{14}\Msol$ with a concentration of $c_{200m}=4.3^{+2.2}_{-1.4}$. In this system a strong lensing analysis of a candidate multiply imaged source is possible.

\section*{Acknowledgements}

This work was supported by SFB-Transregio 33 `The Dark Universe' by the Deutsche Forschungsgemeinschaft (DFG) and the DFG cluster of excellence `Origin and Structure of the Universe'. The authors thank Megan Donahue, August E. Evrard and the anonymous referee for useful comments on the manuscript. VV thanks Bhuvnesh Jain for helpful discussions.

\addcontentsline{toc}{chapter}{Bibliography}
\bibliographystyle{mn2e}
\bibliography{literature}

\label{lastpage}

\end{document}